\documentclass[twoside,12pt]{article}

\usepackage{oldgerm}
\usepackage{amssymb}
\usepackage{amsmath, mathrsfs,psfrag}
\numberwithin{equation}{section}

\newlength{\dinwidth}
\newlength{\dinmargin}
    
\renewcommand{\theequation}{\thesection.\arabic{equation}}

\newtheorem{theorem}{Theorem}[section]
\newtheorem{proposition}[theorem]{Proposition}

\newtheorem{lemma}[theorem]{Lemma}

\newenvironment{proof}{\medskip \noindent {\bf Proof:}}{\hfill
  $\square$ \\[2mm] \indent}

\font\fa=bbm11
\newcommand{\N}{\mbox{\fa N}}   

\newcommand{\Ibb}[1]{ {\rm I\ifmmode\mkern -3.6mu\else\kern -.2em\fi#1}}
\newcommand{\ibb}[1]{\leavevmode\hbox{\kern.3em\vrule
     height 1.2ex depth -.3ex width .2pt\kern-.3em\rm#1}}
\newcommand{\Cl}{{\ibb C}}           
\newcommand{\Rl}{{\Ibb R}}           

\newcommand{\Hil}{\mathcal{H}}
\newcommand{\DD}{\mathcal{D}} 
\newcommand{\W}{\mathcal{W}}    
\newcommand{\Ss}{\mathscr{S}}   
\newcommand{\OO}{\mathcal{O}}   
\newcommand{\PGpo}{\mathcal{P}_+^\uparrow}   

\newcommand{\supp}{\mathrm{supp}\,}
\newcommand{\sh}{\mathrm{sh}}
\newcommand{\ch}{\mathrm{ch}}

\newcommand{\te}{\theta}
\newcommand{\la}{\lambda}
\newcommand{\eps}{\varepsilon}

\newcommand{\fti}{\widetilde{f}}
\newcommand{\fhat}{\hat{f}}
\newcommand{\gti}{\widetilde{g}}
\newcommand{\fbar}{\overline{f}}

\newcommand{\zd}{z^{\dagger}}
\newcommand{\Tu}{\mathscr{T}}
\newcommand{\tp}[1]{^{\otimes #1}}    

\newcommand{\bte}{{\mbox{\boldmath $\theta$}}}
\newcommand{\bze}{{\mbox{\boldmath $\zeta$}}}
\newcommand{\bla}{{\mbox{\boldmath $\lambda$}}}
\newcommand{\bbe}{{\mbox{\boldmath $\beta$}}}
\newcommand{\beps}{{\mbox{\boldmath $\varepsilon$}}}

\newcommand{\sbte}{{\mbox{\footnotesize \boldmath $\theta$}}}

\newcommand{\sbeps}{{\mbox{\footnotesize \boldmath $\varepsilon$}}}

\newcommand{\by}{\mathrm{\bf y}}
\newcommand{\sby}{\mathrm{\footnotesize\bf y}}
\newcommand{\bx}{\mathrm{\bf x}}
\newcommand{\bz}{\mathrm{\bf z}}
\newcommand{\sbx}{\mathrm{\footnotesize\bf x}}

\newcommand{\A}{\mathcal{A}}
\newcommand{\B}{\mathcal{B}}

\newcommand{\bref}[1]{(\ref{#1})}
\renewcommand{\thetheorem}{\arabic{section}.\arabic{theorem}}

\title{Towards the construction of quantum field theories from a
  factorizing S-matrix\footnote{Based on a talk given at the symposium
    ``rigorous quantum field theory'' in the honour of Jacques Bros,
    19.-21.7.2004, Saclay. To appear in the
    conference proceedings.}}

\author{Gandalf Lechner\\
Institut f\"ur Theoretische Physik, Universit\"at G\"ottingen,\\
37077 G\"ottingen, Germany\\
e-mail: {\tt lechner@physik.uni-goe.de}}
\date{February 21, 2005}

\begin{document}

\renewcommand{\theequation}{\thesection.\arabic{equation}}

\maketitle

\begin{abstract}
Starting from a given factorizing S-matrix $S$ in two space-time dimensions, 
we review a novel strategy to rigorously construct quantum field theories 
describing particles whose interaction is governed by $S$. The construction 
procedure is divided into two main steps: Firstly certain semi-local
Wightman fields are introduced by means of Zamolodchikov's
algebra. The second step  consists in proving the existence of local
observables in these models.\\
As a new result, an intermediate step in the existence problem is
taken by proving the modular compactness condition for wedge algebras.
\end{abstract}

\section{Introduction}

In quantum field theory, the rigorous construction of models with
non-trivial interaction is one of the most challenging open
problems. Although collision theory has been established
a long time ago and the calculation of the scattering matrix is well understood, little is known
about the inverse problem, {\em i.e.} the reconstruction of
interacting models from a given S-matrix. The only situation in which
certain steps of such a reconstruction have been carried out is the
class of factorizing S-matrices on two-dimensional Minkowski
space, which correspond to scattering processes in which the particle
number and momenta are conserved. This issue is usually taken up in the framework of the
so-called formfactor program \cite{ffp, BKW, smirnov}, which aims at the construction of local quantum field theories corresponding to factorizing S-matrices by determining expectation values of local 
operators in scattering states. In spite of the interesting results
that have been obtained for many S-matrices, the final
construction of interacting Wightman fields has not been achieved up
to now.\\
\indent In the present paper, we shall review a novel approach to this
construction problem, which has been initiated by Schroer in the last few
years (see also the contribution of B.~Schroer to these conference
proceedings \cite{schroer-rqft}). As it mainly uses the framework of local quantum physics
\cite{araki, haag} instead of Wightman theory, we will term this new approach
``algebraic'' as opposed to the more field theoretic concepts of the
formfactor program. The starting point of the algebraic approach was
Schroer's insight that for the family of factorizing S-matrices
describing a single type of massive, scalar particles, certain field
operators arising from Zamolodchikov's algebra \cite{zamo} (in which
the given factorizing S-matrix $S$ is encoded) can be interpreted as
being localized in wedge-shaped regions of Minkowski space
\cite{schroer1}. Subsequently, the understanding of these wedge-local
fields was deepened in \cite{schroer-wiesbrock} and \cite{gl1}, and
connection to the algebraic formulation of quantum field theory was made by
investigating the von Neumann algebras generated by them
\cite{BuLe}. The construction of the wedge-local fields and their corresponding operator algebras will be
reviewed in section two.\\
The second step of the algebraic program is devoted to exhibiting {\em local}
observables. Compared to the formfactor program, where the aim is an explicit
construction of local field operators, the algebraic
approach focusses on the question of {\em existence} of local
operators, which can be phrased in terms of the aforementioned
wedge algebras. In \cite{BuLe}, the modular nuclearity condition
\cite{nuclearmaps1, nuclearmaps2} for wedge algebras was put forward as a sufficient
condition for the existence of local observables. This condition was
then shown to be fulfilled in specific models in \cite{BuLe, gl2}. These
subjects will be discussed in section three.\\
\indent Whereas the subsequent sections two and three have the character of a review, we will prove a new result in section four, already announced in \cite{BuLe}. In a
specific class of models with factorizing S-matrices, the modular
compactness criterion for wedge algebras will be verified, thus taking
a further step towards proving the existence of local
observables. Regarding the occasion of this conference, we emphasize
the relation of our compactness proof to the work of J.~Bros \cite{bros:compactness} on the 'Haag-Swieca compactness property'
\cite{haag-swieca}. Inspired by his strategy, we
will also employ techniques of complex analysis in section four.\\
The article ends with a short summary in section five.\\

We conclude this
introductory section by stating our assumptions: In the spirit of the inverse scattering approach, our construction begins with the
specification of the particle content of the theory and the S-matrix. We deal here only with
a single species of scalar particles of mass $m>0$. It will be
convenient to parametrize the upper mass shell by the rapidity
$\theta$ via
\begin{eqnarray}
  p(\te)
  &=&
  m\left(
    \begin{array}{c}
      \cosh\te\\
      \sinh\te
    \end{array}
    \right).
\end{eqnarray}
In this variable, the
physical sheet of the complex energy plane is transformed to the horizontal strip
$S(0,\pi) := \{\zeta\in\Cl\,:\,0 < \mathrm{Im}\zeta < \pi\}$.\footnote{More generally, we will use the notation $S(a,b)=\{\zeta\in\Cl\,:\,a < \mathrm{Im}\zeta < b\}$ in the following.}\\
As mentioned
before, the S-matrix is assumed to be of the factorizing type. This
implies that it can be described by means of a single function $S_2$, called the {\em scattering function} in the following,
which is related to two-particle S-matrix elements by
\begin{eqnarray}
  {}_{\mathrm{out}}\langle\te_1,\te_2|\te_1,\te_2\rangle_{\mathrm{in}} &=& S_2(|\te_1-\te_2|)\,,
\end{eqnarray}
and is required to satisfy the following conditions:
\begin{enumerate}
\item $S_2:\overline{S(0,\pi)}\to\Cl$ is continuous and 
  bounded, and analytic on $S(0,\pi)$.
\item For real $\theta$ one has
  \begin{equation}\label{s2-constraints}
    S_2(\te)^{-1} = S_2(-\te) = \overline{S_2(\te)} = S_2(\te+i\pi)\,.
  \end{equation}
\end{enumerate}
By excluding poles of $S_2$ in the strip $S(0,\pi)$, the first
condition characterizes models without bound states. The equations summarized in \bref{s2-constraints} arise from the
requirements of unitarity, crossing symmetry and hermitian analyticity
 for the corresponding S-matrix ({\em cf.}, for example, the review in \cite[p. 46]{castro} and the
references cited there), and put strong
constraints on the possible form of the function $S_2$. Indeed, the general
solution of \bref{s2-constraints} is quite explicitely known
\cite{mitra}. We note here the existence of two particularly simple solutions,
namely $S_2(\te)=\pm 1$, which will be discussed in some detail. A more generic scattering function is given by
\begin{eqnarray}\label{s2g}
  S_2(\te) &=& \frac{1+ig\sinh\te}{1-ig\sinh\te}\,,
\end{eqnarray}
where $g>0$ is some constant. It interpolates between the preceding
solutions in the limit of small and large $g$, respectively. We also
note that products of scattering functions again satisfy \bref{s2-constraints}.

\section{Wedge-local fields}

\subsection{Hilbert space and Zamolodchikov's algebra}

The starting point of the construction of the wedge-local fields is
Zamolodchikov's algebra\footnote{Sometimes also called
  Zamolodchikov-Faddeev algebra} \cite{zamo}, which is a basic ingredient in the context
of factorizing S-matrices. We do not deal with the abstract algebra
here, but rather with a particular representation of it on a conveniently
chosen Hilbert space $\Hil$, which we define first. For details we
refer the reader to \cite{ligmin, gl1}.\\
In view of the assumptions on the particle spectrum made above, a
reasonable choice for the one particle Hilbert space is $\Hil_1 :=
L^2(\Rl,d\te)$, the space of square integrable functions over the upper mass
shell of mass $m$. The $n$-particle space $\Hil_n$ is defined as a particular
subspace of the (unsymmetrized) $n$-fold tensor product $\Hil_1\tp{n}$; namely its
elements are wavefunctions $\psi_n \in L^2(\Rl^n,d^n\te)$ which satisfy the symmetry relations
\begin{eqnarray}\label{psi-sym}
  \!\!\!\psi_n(\te_1,...,\te_{k+1},\te_k,...,\te_n)
  &=&
  S_2(\te_k-\te_{k+1})\cdot \psi_n(\te_1,...,\te_k,\te_{k+1},...,\te_n)
\end{eqnarray}
for $k=1,...,n-1$. Here $S_2$ is the scattering function corresponding
to the S-matrix we are considering. The full Hilbert space of the
theory is
\begin{eqnarray}
  \Hil &:=& \bigoplus_{n=0}^\infty \Hil_n\,,
\end{eqnarray}
where we have put $\Hil_0 := \Cl\cdot\Omega$ to denote the zero particle space
containing the vacuum unit vector $\Omega$. For the special scattering
functions $S_2=1$ and $S_2=-1$, $\Hil$ coincides with
the symmetric and antisymmetric Fock space over $\Hil_1$,
respectively. But in general we deal with a ``twisted'' Fock space
with rapidity dependent symmetry structure.\\
On $\Hil$ we have a positive energy representation $U$ of the proper
orthochronous Poincar\'e group $\PGpo$. If $(x,\la)\in\PGpo$ denotes
the transformation consisting of a boost with rapidity parameter $\la\in\Rl$
and a subsequent translation along $x\in\Rl^2$, $U(x,\la)$ is defined
as, $\psi_n\in\Hil_n$,
\begin{eqnarray}\label{U}
  (U(x,\la)\psi_n)(\te_1,...,\te_n)
  &=&
  e^{i\sum_{k=1}^n p(\te_k)x} \cdot \psi_n(\te_1-\la,...,\te_n-\la)\,.
\end{eqnarray}
In the following, we will also use the shorthand notation $U(x):=U(x,0)$ for pure translations.\\
The creation and annihilation operators
familiar from symmetric and antisymmetric Fock space have their
counterparts on $\Hil$. These operator valued distributions will be
denoted $z(\te)$ and $\zd(\te)=z(\te)^*$, and are defined by
\begin{eqnarray}
  (z(\te)\psi_n)(\te_1,...,\te_n) &=& \sqrt{n}\cdot \psi_n(\te,\te_1,...,\te_n)
\end{eqnarray}
and by taking the adjoint on $\Hil$. This definition yields a representation of Zamolodchikov's algebra on
$\Hil$, {\em i.e.} $z(\te)$, $\zd(\te)$ satisfy the exchange relations
\begin{eqnarray}
  \zd(\te_1)\zd(\te_2) &=& S_2(\te_1-\te_2)\zd(\te_2)\zd(\te_1)\,,\\
  z(\te_1)\zd(\te_2)   &=& S_2(\te_2-\te_1)\zd(\te_2)z(\te_1) + \delta(\te_1-\te_2)\cdot 1\,.
\end{eqnarray}
We will also write $z(\psi)=\int d\te\, z(\te)\psi(\te)$,
$\zd(\psi)=\int d\te\, \zd(\te)\psi(\te)$, for wave functions
$\psi\in\Hil_1$. Note that with these conventions, $z(\psi)^*=\zd(\overline{\psi})$.

\subsection{Wedge locality}

With the help of the creation and annihilation operators $\zd(.)$ and
$z(.)$, a scalar quantum field $\phi$ can be defined on (a dense
domain in) $\Hil$ in a manner analogous to the definition of the free
field on symmetric Fock space. For $f\in\Ss(\Rl^2)$, we consider the
restrictions of the Fourier transform of this function to the upper
and lower mass shell, parametrized by the rapidity:
\begin{eqnarray}
  f^\pm(\te) &:=& \frac{1}{2\pi}\int d^2x\,f(x)e^{\pm i p(\te)x}\,,
\end{eqnarray}
and set 
\begin{eqnarray}\label{def:phi}
  \phi(f) &:=& \zd(f^+) + z(f^-)\,,
\end{eqnarray}
which is a well-defined operator on the dense subspace
$\DD\subset\Hil$ of vectors of finite particle number. In the case of
the scattering function $S_2=1$,
this definition yields the well-known free scalar field. But also
for different scattering functions, $\phi$ has many properties in common with a free field.
\begin{proposition}{\bf \cite{gl1}}
  The field operator $\phi(f)$ has the following properties:
  \begin{enumerate}
  \item $\phi(f)$ is defined on $\DD$ and leaves this space invariant.
  \item For $\psi\in\DD$ one has
    \begin{equation}
      \phi(f)^*\psi = \phi(\fbar)\psi.
    \end{equation}
    All vectors in $\DD$ are entire analytic for
    $\phi(f)$. If $f\in\Ss(\Rl^2)$ is real, $\phi(f)$ is
    essentially self-adjoint on $\DD$.
    \item $\phi$ is a solution of the Klein-Gordon equation: For every
      $f\in\Ss(\Rl^2)$, $\psi\in\DD$ one has 
      \begin{eqnarray}
        \phi((\Box+m^2)f)\psi = 0\,.
      \end{eqnarray}
  \item $\phi(f)$ transforms covariantly under the representation
    $U$ of $\PGpo$, cf. (\ref{U}):
    \begin{equation}\label{phi_cov}
      U(g)\phi(f)U(g)^{-1} = \phi(f_g),\quad f_g(x)=f(g^{-1}x),\quad
      g\in\PGpo\,.
    \end{equation}
  \item The vacuum $\Omega$ is locally cyclic for the field
    $\phi$. More precisely, given any open subset $\OO\subset\Rl^2$, the
    subspace 
    \begin{eqnarray}
      D_\OO := \mathrm{span}\{\phi(f_1)\cdots\phi(f_n)\Omega
    \,:\, f_k \in \Ss(\OO),\,n\in\N_0\}
    \end{eqnarray}
    is dense in $\Hil$.
  \end{enumerate}
\end{proposition}

In spite of these pleasant properties of the field operator, a simple
calculation shows that $\phi$ is local if and only if $S_2=1$. As
locality is one of the fundamental principles in quantum field theory,
the generically non-local field operators $\phi(f)$ cannot be
interpreted as the basic physical fields of our model, but rather as
an auxiliary tool in the construction of the theory: They are polarization-free generators in the sense of \cite{pfgs}.\\
To clarify the
role of the field $\phi$, we consider subsets $W$ of $\Rl^2$ called {\em wedges}, which are the Poincar\'e transforms of the so-called left wedge
\begin{eqnarray}
  W_L &:=& \{x\in\Rl^2\,:\,|x_0|+x_1<0\}\,.
\end{eqnarray}
As $W_L$ is invariant under boosts, any wedge has the form  $W=W_L+x$
or $W=W_R+x$ for some $x\in\Rl^2$, where $W_R=-W_L=W_L'$ is the right wedge. The set of
wedges will be denoted by $\W$.\\
Following Schroer and Wiesbrock \cite{schroer-wiesbrock}, we address
the question whether it is possible to consistently interpret the
field $\phi$ as being localized in a wedge region, say in $W_L$ for
the sake of concreteness. Put differently, we take
\begin{eqnarray}\label{awl}
  \A(W_L) &:=& \{e^{i\phi(f)}\,:\,f\in\Ss_\Rl(W_L)\}''
\end{eqnarray}
as the von Neumann algebra generated by the observables in
$W_L$ and look for a map $\W\ni W\longmapsto\A(W)$ of wedge regions to
von Neumann subalgebras of $\B(\Hil)$ such that $\A(W_L)$ is given by
\bref{awl} and the following standard properties \cite{araki, haag} hold: ($W,W_1,W_2\in\W$)
\begin{enumerate}
\item $\A(W_1)\subset\A(W_2)\;$ for $\;W_1\subset W_2$ (Isotony)
\item $U(g)\A(W)U(g)^*=\A(gW)$, $g\in\PGpo$ (Covariance)
\item $\A(W')\subset\A(W)'$ (Wedge-Locality)
\item $\Omega$ is cyclic for each $\A(W)$ (Reeh-Schlieder property)
\end{enumerate}
Such a map $W\longmapsto\A(W)$ will be called a {\em wedge-local covariant net}. Within the present context one obtains a net by setting
\begin{eqnarray}
  \A(W_R) &:=& \A(W_L)',\label{aw2}\\
  \A(W+x) &:=& U(x)\A(W)U(x)^*,\qquad x\in\Rl^2,\;\, W\in\W,\label{aw3}
\end{eqnarray}
where the prime denotes taking the commutant in $\B(\Hil)$. Whereas
the first three properties (isotony, covariance, and wedge locality)
follow immediately from the definitions (\ref{awl}-\ref{aw3}), the
cyclicity of $\Omega$ for $\A(W_R)$ is not so obvious -- Proving it is
equivalent to showing that $\phi$ can be interpreted as being localized in $W_L$.
\begin{proposition}{\bf\cite{gl1,BuLe}}\\
The correspondence $W\longmapsto\A(W)$ defined in
  (\ref{awl},\ref{aw2},\ref{aw3}) is a wedge-local covariant
  net. In particular, $\Omega$ is cyclic and seperating for each $\A(W)$,
  $W\in\W$. Moreover, $\A(W')=\A(W)'$, i.e. wedge duality holds.
\end{proposition}
The cyclicity of $\Omega$ can be proven by considering the antilinear operator $J$,
\begin{eqnarray}
  (J\psi)_n(\te_1,...,\te_n)
  &:=&
  \overline{\psi_n(\te_n,...,\te_1)}\,,
\end{eqnarray}
which can be adjoined to the representation $U$ to obtain a
representation of the proper Poincar\'e group $\mathcal{P}_+$. More
importantly, it gives rise to a second field $\phi'$,
\begin{eqnarray}
  \phi'(f) &:=& J\phi(f^j)J,\qquad\qquad f^j(x) := \overline{f(-x)}\,.
\end{eqnarray}
The ``reflected field'' $\phi'(f)$, can be shown
to commute with $\phi(g)$, in the sense that their
associated unitary groups commute, whenever $\supp f+W_R$ is spacelike
seperated from $\supp g+W_L$. As the vacuum is cyclic for
$\phi'$ as well, the cyclicity of $\Omega$ for all wedge algebras then
follows.\\
In the next section, the modular data associated to $(\A(W_L),\Omega)$
will become important. As the $J$ maps $\A(W_L)$ onto $\A(W_R)$, it
can be shown to coincide with the modular conjugation of this couple. The
modular group $\Delta_L^{i\la}$ of $(\A(W_L),\Omega)$ acts as expected from the Bisognano-Wichmann theorem \cite{BiWi, borchers, mund}:

\begin{proposition}\label{prop-W}{\bf \cite{BuLe}}\\
  The modular group and  conjugation of $(\A(W_L),\Omega)$ are given by $\Delta_L^{i\la} = U(0,2\pi\la)$ and $J$, respectively.
\end{proposition}

Before entering into the discussion of local observables in these
models, we mention that it is possible to calculate the two-particle
scattering states with the help of the wedge-local fields $\phi$ and
$\phi'$, since left and right wedges can be causally seperated by
translation. Because of wedge-locality, it turns out that the
particles are ``Bosons''. It is then possible to determine the
two-particle S-matrix, which is 
the ``right'' one, {\em i.e.} the one associated to the scattering
function $S_2$ we started with \cite{gl1}. The construction of the
wedge algebras thus leads to a (wedge-local) quantum field theory of
particles whose interaction is described by $S_2$. 

\section{Existence of local observables}

\subsection{Observables localized in a double cone}

Having constructed a wedge-local quantum theory with the correct
two-particle scattering states, the next
step is to exhibit observables
localized in {\em bounded} space-time regions. Typical examples of such
regions are double cones, which in two dimensions can always be realized as 
intersections of two opposite wedges. To fix ideas, consider the
double cone
\begin{eqnarray}
  \OO_x &:=& W_R \cap (W_L+x),\qquad x\in W_R\,.
\end{eqnarray}
An operator $A$ representing an observable localized in $\OO_x$ has to
commute with any observable localized in ${\OO_x}' = W_L \cup (W_R+x)$
because of Einstein causality. Any such $A$ is therefore an element of the
algebra
\begin{eqnarray}\label{ao}
  \A(\OO_x)
  &:=&
  \left(\A(W_L)\vee\A(W_R+x)\right)'
  =
  \A(W_R) \cap \A(W_L+x)\,,
\end{eqnarray}
the relative commutant of $\A(W_R+x)$ in $\A(W_R)$. We will adopt
\bref{ao} as the definition of the algebra generated by the
observables localized in $\OO_x$ in our model. The algebras associated to translated double cones are then fixed by covariance.\\
The net $\OO\longmapsto\A(\OO)$ of double cone algebras arising in
this manner inherits the basic properties of isotony, covariance and
locality from the corresponding features of the wedge net, as can be
verified in a straightforward manner. But without further information
on the structure of the wedge algebras, it is not clear whether the
relative commutants \bref{ao} are non-trivial. As a physical theory should describe local observables, we would like to rule out the pathological cases in which $\A(\OO_x)=\Cl\cdot 1$.\\
\indent In \cite{schroer-wiesbrock}, a method to construct explicitely
non-trivial operators localized in $\OO_x$ has been proposed. However,
this procedure faces substantial difficulties related to the
convergence of certain ``perturbation'' series. We will concentrate
here on an existence proof without trying to give concrete expressions for local operators.

\subsection{Split property and modular nuclearity}

The basic idea in the approach to the existence problem proposed in
\cite{BuLe} is the observation that the non-triviality of the relative
commutant $\A(\OO_x)$ \bref{ao} can be established if the net of wedge
algebras $\A(W)$ has the {\em split property}, {\em i.e} if to each
$x\in W_R$ there is a type I$_\infty$ von Neumman factor $\mathcal{N}_x$ such that
\begin{eqnarray}\label{split}
  \A(W_R+x) \subset \mathcal{N}_x \subset \A(W_R)\,.
\end{eqnarray}
In this case, the observables localized in $W_L$ and $W_R+x$ satisfy a
strong form of statistical independence (for a review, see \cite{summers:statistical}),
and the algebraic structure of $\A(\OO_x)$ is completely
fixed. According to a result of Longo \cite{longo}, the wedge algebras
$\A(W)$  are type III$_1$ von Neumann factors in the present context. Using this information and the split assumption \bref{split}, one can establish the unitary
equivalence \cite{dopl, BuLe}
\begin{eqnarray}
  \A(\OO_x)\cong\A(W_R)\otimes\A(W_L+x).
\end{eqnarray}
Thus the split property for the wedge algebras implies that the local
algebras $\A(\OO_x)$ are of type III as well, and in particular non-trivial.\\
In the following, the split
property will be used as a sufficient condition for the non-triviality
of the relative commutants \bref{ao}\footnote{Note that in two
  space-time dimensions the split property for wedges is a reasonable
  assumption since Araki's argument \cite{bu:productstates} to the effect
  that inclusions of wedge algebras
cannot be split does not apply here because of the missing translation
invariance along the edge of the wedge.}. However, as the existence of an
interpolating type I factor \bref{split} is difficult to establish
directly, one needs another condition, implying the split property, which is
better manageable in concrete models. In the literature different
``nuclearity'' criteria for the split property have been discussed,
the one which is relevant in our context being introduced in
\cite{nuclearmaps1}. As these criteria involve the notions of nuclear maps, we recall that a bounded operator between two Banach spaces is called {\em nuclear} if it can be expanded into a norm convergent series of rank one operators \cite{pietsch}. Note that a nuclear map is in particular compact.\\
To formulate the relevant condition for the split property, we denote by
$J,\Delta$ the modular involution and modular operator of the pair
$(\A(W_R),\Omega)$, respectively\footnote{$\Delta$ is connected to
  the earlier discussed modular operator of the left wedge by $\Delta=\Delta_L^{-1}$.}, and introduce the maps
\begin{eqnarray}\label{def:Xi}
  \begin{array}{rcl}
  \Xi(x) : \A(W_R) &\longrightarrow& \Hil,\\
  \vspace*{-3mm}\\
  A &\longmapsto& \Delta^{1/4}U(x)A\Omega\,.
  \end{array}
\end{eqnarray}
Using modular theory, one easily finds
\begin{eqnarray*}
  \!\!\!\!\!\!\|\Xi(x)A\|^2 &=& \langle U(x)A\Omega,\Delta^{1/2}U(x)A\Omega\rangle
  = \langle U(x)A\Omega,JU(x)A^*\Omega\rangle \leq \|A\|^2,
\end{eqnarray*}
{\em i.e.} $\Xi(x)$ is a bounded map with $\|\Xi(x)\|\leq 1$ for any
$x\in \overline{W_R}$. Based on results of \cite{nuclearmaps1}, the following ``modular nuclearity condition'' has been discussed in \cite{BuLe}.

\begin{proposition}\label{prop-nuc}{\bf\cite{BuLe}}\\
  If $\Xi(x)$ \bref{def:Xi} is nuclear, the inclusion
  $\A(W_R+x)\subset\A(W_R)$ is split and the local
  algebra $\A(\OO_x)$ \bref{ao} is isomorphic to the unique
  hyperfinite type III$_1$ factor.
\end{proposition}
Proposition \ref{prop-nuc} is used as a sufficient condition for the
non-triviality of the local algebras in the model theories presented in section two. 
If the scattering function is constant, $S_2(\te)=\pm 1$, the
structure of Zamolodchikov's algebra simplifies to the CCR- and
CAR-algebra, respectively, and the estimates needed for the nuclearity
proof of $\Xi(x)$ are fully under control. We have
\begin{proposition}\label{prop:pm1}{\bf\cite{BuLe, gl2}}\\
In the models corresponding to the constant scattering functions $S_2(\te)= 1$
and $S_2(\te)=-1$, the maps $\Xi(x)$ \bref{def:Xi} are nuclear, $x\in W_R$. In particular,
the split property for wedges holds and all double cone algebras
\bref{ao} contain non-trivial observables.
\end{proposition}
The proof for the case $S_2(\te)=1$ can be found in \cite{BuLe}, where
previous results obtained in $\cite{BuJu,BuWi}$ have been used. For
the case $S_2(\te)=-1$, see \cite{gl2}. In these articles, one also finds
explicit bounds on the nuclear norms $\|\Xi(x)\|_1$. The case $S_2(\te)=1$ gives just the free scalar field in two dimensions, and the model corresponding to the scattering function $S_2(\te)=-1$ is related to the scaling limit of the Ising model (see
\cite{BKW} and the references cited there).\\
Although the existence of local observables is well known in free
field theory, the check of the modular nuclearity condition in the
case $S_2(\te)=1$ was an important test for its value in the discussion of models with non-constant scattering functions. In view of Proposition \ref{prop:pm1} and the earlier mentioned 
fact that $S_2(\te)=\pm 1$ may be considered as the ``limiting cases'' of typical non-constant scattering functions, we conjecture that $\Xi(x)$ 
is nuclear in the family of models considered in section two.\\
\indent It was shown in \cite{nuclearmaps1} that whereas the nuclearity of
$\Xi(x)$ is sufficient for the split property, the {\em compactness}
of $\Xi(x)$ is a necessary condition for split. In the next section,
our conjecture about the nuclearity of $\Xi(x)$ will be further
substantiated by proving the compactness of this map in a wide class of models with certain typical scattering functions.

\section{Modular Compactness for Wedge Algebras}

In this section we concentrate on the model corresponding to the
scattering function \bref{s2g} with arbitrary constant $g>0$, or a
finite product of such functions with different values of $g$. In view
of the general solution \cite{mitra} of the constraining equations
\bref{s2-constraints} for $S_2$, this is a typical example of a
non-constant scattering function. The aim of this section is the proof
of the following Proposition.

\begin{proposition}\label{prop-compact}
  Consider the model theory corresponding to the scattering function
  \begin{eqnarray}\label{s2gg}
    S_2(\te) &:=& \prod_{r=1}^R \frac{1+ig_r\sinh\te}{1-ig_r\sinh\te}, 
  \end{eqnarray}
where $R<\infty$ and  $g_1,...,g_R > 0$. The maps $\Xi(x)$ are compact, $x\in W_R$.
\end{proposition}
Before explaining our strategy of the proof, we make a few remarks about the maps $\Xi(x)$ and introduce some notation.\\
First note that it
is sufficient to consider the maps $\Xi(0,s)$ corresponding to wedge inclusions of the type
\begin{eqnarray}\label{wsinc}
  W_R+(0,s) &\subset& W_R,\qquad s>0\,.
\end{eqnarray}
As $W_R$ is stable under boosts, a more general inclusion
$W_R+x\subset W_R$, $x\in W_R$, of right wedges can be transformed to
\bref{wsinc} by a velocity transformation with
appropriately chosen rapidity parameter $\la$. Using the fact that
the boosts commute with the modular operator, one easily shows $\Xi(x) =
U(0,-\la)\Xi(0,s)\mathrm{Ad}U(0,\la)$, where
$s=(x_1^2-x_0^2)^{1/2}>0$. Hence $\Xi(x)$, $x\in W_R$, is nuclear
(compact) if and only if $\Xi(0,s)$, $s>0$, is nuclear (compact). In
the case of nuclear maps, $\|\Xi(x)\|_1 =
\|\Xi(0,(x_1^2-x_0^2)^{1/2})\|_1$. For this reason, we will consider in
the following only inclusions of the form \bref{wsinc}, and use the shorthand notation $\Xi(s):=\Xi(0,s)$.\\
\indent It will be useful to consider, as a generalization of \bref{def:Xi}, the maps
\begin{eqnarray}
  \Xi^\alpha(s)
  \A(W_R) &\longrightarrow& \Hil\,,\quad
  \Xi^\alpha(s)A :=   \Delta^{\alpha}U(s)A\Omega\,,\qquad 0\leq\alpha\leq\tfrac{1}{2},
\end{eqnarray}
and we adopt the convention to suppress the upper index for the
``canonical'' value $\alpha=\frac{1}{4}$. Furthermore, we introduce the $n$-particle
projections\footnote{Note that the modular operator $\Delta$ can be
  restricted to the $n$-particle space $\Hil_n$.}
\begin{eqnarray}\label{xi-n-a}
  \Xi_n^\alpha(s) 
  &:=&
  P_n\Xi^\alpha(s)\,,
\end{eqnarray}
where $P_n\in\B(\Hil)$
denotes the orthogonal projection onto the $n$-particle subspace $\Hil_n$.\\

In the proof of Proposition \ref{prop-compact}, the maps $\Xi_n(s)$ will be shown to be nuclear by estimating the ``size'' of their images in $\Hil_n$. This is achieved by exploiting certain analytic properties of the $n$-particle rapidity wavefunctions 
\begin{eqnarray}\label{wave}
  \psi_n^s &:=& P_n U(0,s) A \Omega\,,\qquad A\in\A(W_R),
\end{eqnarray}
which are considered as elements of $L^2(\Rl^n,d^n\bte)$ and
constitute our main objects of interest in the following. (For a
different compactness proof based on the techniques of complex
analysis, see \cite{bros:compactness}.)\\
From modular theory and the Bisognano-Wichmann property (as specified
in Proposition \ref{prop-W}) it is known that
\begin{eqnarray}
  \la \longmapsto (\Delta^{-i\la/2\pi} \psi_n^s)(\te_1,...,\te_n) = \psi_n^s(\te_1-\la,...,\te_n-\la)
\end{eqnarray}
is a strongly analytic function in the strip $S(0,\pi)$, and $\|\Delta^{\la/2\pi}\psi_n^s\|\leq\|A\|$, $0\leq\la\leq\pi$. In particular, the vectors in the image of $\Xi_n(s)$ are given by the functions
\begin{eqnarray}
  (\te_1,...,\te_n) \longmapsto 
  (\Delta^{1/4}\psi_n^s)(\te_1,...,\te_n)
  &=&
  \psi_n^s(\te_1-\tfrac{i\pi}{2},...,\te_n-\tfrac{i\pi}{2}),
\end{eqnarray}
which have an analytic continuation  (in the sense of distributions)
in the ``center of mass rapidity'' $n^{-1}\cdot(\te_1+...+\te_n)$ to
the strip $S(-\frac{\pi}{2},\frac{\pi}{2})$. Furthermore, the
$L^2$-bound of the continuation is uniformly bounded over this strip and the
convergence to the boundary values is valid in the norm topology of $\Hil_n$.\\
The main idea in the proof of Proposition \ref{prop-compact} consists
in the observation that in the models at hand, the wavefunctions
\bref{wave} enjoy considerably stronger analytic properties, namely
they are holomorphic, as functions of $n$ complex variables, in a
certain tube domain. More precisely, we find:
\begin{lemma}\label{lemma-n-rough}
  Let $A\in\A(W_R)$, $s>0$, $n\in\N_0$ and $\psi_n^s$ as in \bref{wave}. There exists a constant $\alpha > 0$ (independent of $A$ and $s$, but dependent on $n$) such that
  \begin{itemize}
  \item[a)]$\Delta^\alpha\psi_n^s$ is analytic in the tube 
    \begin{eqnarray}
      \mathcal{T}_n(\alpha)
      &:=&
      \{\bze\in\Cl^n \,:\, -2\pi\alpha < \mathrm{Im}(\zeta_k) < 2\pi\alpha, \; k=1,...,n\}.      
    \end{eqnarray}
  \item[b)] For any $\bla\in \;]\!-2\pi\alpha,2\pi\alpha\,[\,^{\times n}$, 
    \begin{eqnarray}
      \Rl^n \ni \bte \longmapsto (\Delta^\alpha\psi_n^s)(\bte+i\bla)
    \end{eqnarray}
    is in $L^2(\Rl^n,d^n\bte)$, with norm bounded by
    $K\cdot\|A\|$, where $K$ depends on $\alpha$, $s$ and $n$, but is
    independent of $A$ and $\bla$. Moreover, $\lim\limits_{|\sbte|\to\infty}|\psi_n^s(\bte+i\bla)|=0$.
  \item[c)] $(\Delta^\alpha\psi_n^s)(\bze)$ converges strongly
    to its boundary values at {\em Im}$(\zeta_k)=\pm2\pi\alpha$.
  \end{itemize}
\end{lemma}
Here we introduced the convention to denote vectors in $\Cl^n$ or
$\Rl^n$ by bold face letters $\bze,\bla,\bte$, and their components by
$\zeta_k,\la_k,\te_k$, $k=1,...,n$. Note in particular that by
considering the limit Im$(\zeta_k)\to 2\pi\alpha$, $k=1,...,n$, the
wavefunctions $\psi_n^s$ \bref{wave} are recovered as a (strong) boundary value of
the analytic function $\Delta^\alpha\psi_n^s$.\\
The constants $\alpha$ and $K$ appearing in Lemma \ref{lemma-n-rough}
specify the size of the tube $\mathcal{T}_n(\alpha)$ in which $\Delta^\alpha\psi_n^s$ is
analytic and its bound in that region, respectively. They depend on
the scattering function $S_2$ \bref{s2gg} under consideration, {\em
  i.e.} on the parameters $g_1,...,g_R$. This dependence will
be made explicit in the proof of Lemma \ref{lemma-n-rough}, which is
based on wedge locality and the symmetry properties \bref{psi-sym} of $n$-particle functions. However, it will require the discussion of some technical points. We therefore postpone it and and rather begin by showing how Lemma \ref{lemma-n-rough} can be used to derive estimates on the nuclear norms of $\Xi_n(s)$.
\begin{lemma}\label{lemma-nuc-n}
  The maps $\Xi_n(s)$ are nuclear, $s>0$.
\end{lemma}
\begin{proof}
In view of the definition of the translations \bref{U}, we have, $\bte\in\Rl^n$, 
\begin{eqnarray}\label{tranz}
  (\Delta^\alpha \psi_n^s)(\bte)
  &=&
  (\Delta^\alpha U(0,\tfrac{s}{2})\psi_n^{\frac{s}{2}})(\bte)\nonumber\\
  &=&
  e^{-im\frac{s}{2}\sum_{k=1}^n \sh(\te_k-2\pi\alpha\, i)}(\Delta^\alpha\psi_n^{\frac{s}{2}})(\bte)\nonumber\\
   &=&
   \prod_{k=1}^n e^{-i\frac{ms}{2}\cos(2\pi\alpha)\,\sh\te_k}e^{-\frac{ms}{2}\sin(2\pi\alpha)\,\ch\te_k}\cdot(\Delta^\alpha\psi_n^{\frac{s}{2}})(\bte).
 \end{eqnarray}
 The strongly decreasing factor appearing here allows us to conclude
 the nuclearity of $\Xi_n(s)$ by application of Cauchy's integral
 formula: Let $\bte\in\Rl^n$ and consider a polydisc
 $\mathscr{D}\subset\mathcal{T}_n(\alpha)$ with
 $\bte\in\mathscr{D}$. As $\Delta^\alpha \psi_n^s$ is holomorphic
 in $\mathcal{T}_n(\alpha)$,
  \begin{eqnarray*}
    (\Delta^{\alpha}\psi_n^s)(\bte)
    &=&
    \frac{1}{(2\pi i)^n}\oint_{\partial\mathscr{D}}d^n\bze'\,
    \frac{(\Delta^\alpha\psi_n^s)(\bze')}{\prod_{k=1}^n(\zeta_k'-\te_k)}\, .
  \end{eqnarray*}
  Because of the decrease properties of $\Delta^\alpha\psi_n^s$ in
  real directions, and the good convergence to its boundary values, as
  specified in Lemma \ref{lemma-n-rough} b),c), we can deform the contour of integration to the boundary of $\mathcal{T}_n(\alpha)$ and get
  \begin{eqnarray}\label{int-form}
    (\Delta^{\alpha}\psi_n^s)(\bte)
    &=&
    \frac{1}{(2\pi i)^n}\sum_{\beps}\int\limits_{\Rl^n}d^n\bte'
    \prod_{k=1}^n\frac{\eps_k}{(\te'_k-\te_k-i\cdot 2\pi\alpha\eps_k)}
    \label{int-eqn}\\
    &&\quad\qquad\qquad\qquad\qquad\times\;\, (\Delta^\alpha\psi_n^s)(\te_1'-i\cdot 2\pi\alpha \eps_1,...,\te_n'-i\cdot 2\pi\alpha \eps_n)\nonumber .
  \end{eqnarray}
  The summation extends over the $2^n$ terms parametrized by
  $\eps_1,...,\eps_n = \pm 1$. Consider the integral operators
  $T^\pm_{s,\alpha}\in\B(L^2(\Rl,d\te))$ which are defined by their kernels
  \begin{eqnarray}\label{int-kernelss}
    T_{s,\alpha}^\pm(\te,\te')
    &:=&  
    \pm\frac{1}{i\pi}\frac{e^{-\frac{ms}{2}\sin(2\pi\alpha)\,\ch\te}}{\te'-\te\mp 2\pi i\alpha},
  \end{eqnarray}
and the unitary operator $M_{s,\alpha}$ multiplying with
$e^{-i\frac{ms}{2}\cos(2\pi\alpha)\sh\te}$. With these definitions, inserting \bref{tranz} into \bref{int-form} yields 
\begin{eqnarray}
  \Xi^\alpha_n(s)A = 
\Delta^{\alpha}\psi_n^s
  &=&
  2^{-n}\sum_{\beps}
  {M_{s,\alpha}}^{\otimes n}
  (T_{s,\alpha}^{\eps_1}\otimes ... \otimes T_{s,\alpha}^{\eps_n})
  \Delta^\alpha\psi^{\frac{s}{2}}_{n,\sbeps}\,,\label{alg-eqn}
\end{eqnarray}
where we used the shorthand notation
$\psi^{\frac{s}{2}}_{n,\sbeps}(\bte)=\psi_n^{\frac{s}{2}}(\bte-2\pi\alpha
i \cdot\beps)$. It is now important to note that
$T_{s,\alpha}^\pm$ are trace class operators on $L^2(\Rl,d\te)$
\cite{BuLe}. (This can be seen by Fourier transforming the kernels
\bref{int-kernelss} in the variable $\te'$, and the applying standard
techniques to split the resulting operators into a product of two
Hilbert Schmidt maps \cite[Thm. XI.21]{simon-reed-3}.) Note also that
$T^+_{s,\alpha}$ and $T^-_{s,\alpha}$ are unitarily equivalent, the
equivalence being given by the operator $V$,
$(V\psi)(\te):=\psi(-\te)$. Hence they have the same trace norm $\|T^+_{s,\alpha}\|_1$.\\
According to Lemma \ref{lemma-n-rough} b),
$\|\psi_{n,\sbeps}^{\frac{s}{2}}\|\leq K\|A\|$ for each
$\beps$ occuring in the above summation. Put differently, $A\longmapsto \Delta^\alpha \psi_{n,\sbeps}^{\frac{s}{2}}$ is bounded as a linear map between the Banach spaces $\A(W_R)$ and $\Hil_n$, with norm dominated by $K$. As $M_{s,\alpha}$ is unitary, this implies that $\Xi_n^\alpha(s)$ is a nuclear map, and as a crude bound on its nuclear norm we have 
\begin{eqnarray}\label{crude-bound}
  \|\Xi_n^\alpha(s)\|_1 &\leq& K\,\|T_{s,\alpha}^+\|_1^n\,.
\end{eqnarray}
To proceed from this nuclearity result to the statement in Lemma \ref{lemma-nuc-n}, the power of the modular operator needs to be adjusted from $\alpha$ to
$\frac{1}{4}$. This can be achieved as in \cite[Cor. 3.4]{nuclearmaps1}: Let
$A\in\A(W_R)$. From modular theory we know 
\begin{eqnarray}
  \Xi_n^{\alpha}(s)A
  &=& 
  P_n\Delta^{\alpha}U(0,s)A\Omega
  =
  P_n\Delta^{\alpha}J\Delta^{1/2}U(0,s)A^*\Omega\\
  &=&
  JP_n\Delta^{1/2-\alpha}U(0,s)A^*\Omega
  =
  J\Xi_n^{1/2-\alpha}(s)A^*\,,
\end{eqnarray}
where we used the fact that $\Delta$ and $J$ commute with $P_n$. Hence
$\Xi_n^{1/2-\alpha}(s)$ is nuclear, too, and $\|\Xi_n^{1/2-\alpha}(s)\|_1  =
\|\Xi_n^\alpha(s)\|_1$. Since $X:= \Delta^{1/4}(\Delta^\alpha +
\Delta^{1/2-\alpha})^{-1}$ is a bounded operator with norm $\|X\|\leq
\frac{1}{2}$, it follows from 
\begin{eqnarray}
  \Xi_n(s)
  &=&
  X\Big(\Xi^\alpha_n(s)+\Xi^{1/2-\alpha}_n(s)\Big)
\end{eqnarray}
that $\Xi_n(s)$ is nuclear, with nuclear norm bounded by
\begin{eqnarray}
  \|\Xi_n(s)\|_1
  \leq
  \|X\|\left(\|\Xi^\alpha_n(s)\|_1 + \|\Xi^{1/2-\alpha}_n(s)\|_1\right)
  \leq
  \|\Xi^\alpha_n(s)\|_1\,.
\end{eqnarray}
This completes the proof of the Lemma.
\end{proof}

The proof of Proposition \ref{prop-compact} now follows as a short
corollary of Lemma \ref{lemma-nuc-n} \cite{BuLe}. Let $A\in\A(W_R)$. Since
\begin{eqnarray*}
  (\Xi_n(s)A)(\bte)
  =
  \prod_{k=1}^n
  e^{-ms\cosh\te_k}\cdot (\Xi_n(0)A)(\bte)
  \leq 
  e^{-msn}\cdot (\Xi_n(0)A)(\bte)
\end{eqnarray*}
and $\|\Xi_n(0)\|\leq 1$, we have $\|\Xi_n(s)\| \leq e^{-msn}$, and
hence the series $\sum_{n=0}^\infty \Xi_n(s)$ converges in the norm
topology of $\B(\A(W_R),\Hil)$ to $\Xi(s)$. But as nuclear maps, the $\Xi_n(s)$ are in particular compact, and the set of compact
operators between two Banach spaces is norm closed
\cite[Thm. VI.12]{simon-reed-1}. Thus $\Xi(s)$ is compact, too.{\hfill $\square$ \\[2mm]}
\noindent To accomplish the existence proof for local observables, one has to show that the map $\Xi(s)$ is not only compact, but also nuclear. This amounts to proving that the series $\sum_n\Xi_n(s)$ converges not only in the operator norm $\|\cdot\|$, but also in the nuclear norm $\|\cdot\|_1$, {\em i.e.} better bounds than \bref{crude-bound} on $\|\Xi_n(s)\|_1$ are required. Such a refined analysis will be presented elsewhere.\\

After having shown how analytic properties of the wavefunctions
$\psi_n^s$ lead to the nuclearity of $\Xi_n(s)$, it remains to derive
these properties, {\em i.e.} prove Lemma \ref{lemma-n-rough}. This
proof will be devided into three steps: First, we consider the
dependence of $\psi_n^s$ on its first variable $\te_1$ only, and
derive analytic properties of $\te_1\mapsto\psi_n^s(\te_1,...,\te_n)$
by using the localization of $A$ in $W_R$ (Lemmata \ref{lemma-C} and
\ref{lemma-h}). This is a kind of one-particle analysis, and the form
of the scattering function $S_2$ does not matter here. In a second
step, the symmetry \bref{psi-sym} is exploited to transfer these
results to the other variables $\te_2,...,\te_n$. Here two important
properties (\ref{s-kappa}, \ref{uniform-s-bound}) of $S_2$ enter. Finally, the $n$-variable analyticity claimed in Lemma \ref{lemma-n-rough} is established by using the Malgrange-Zerner (``flat tube'') theorem ({\em cf.}, for example, \cite{epstein}).\\

To exploit the localization of $A$, it is useful to consider the time
zero fields $\varphi,\pi$ of the ``left localized'' field $\phi$
\bref{def:phi} and study expectation value functionals of the
commutator of these fields with $A$. In rapidity space, these
functionals give rise to certain analytic functions (Lemma
\ref{lemma-C}). Their relations to the wavefunctions $\psi_n^s$ are
explained in Lemma \ref{lemma-h}.\\
The formal definitions $\varphi(x)=\phi(0,x)$, $\pi(x) = (\partial_0\phi)(0,x)$, $x\in\Rl$, can be rephrased as ($f\in\Ss(\Rl)$)
\begin{eqnarray}
  \fhat(\te) &:=& \fti(m\sinh\te),\qquad \fhat_-(\te) := \fhat(-\te),\\
  \varphi(f) &:=& \zd(\fhat) + z(\fhat_-),\label{def-phi}\\
  \pi(f)     &:=& i(\zd(\omega\fhat) - z(\omega\fhat_-))\label{def-pi}.
\end{eqnarray}
Here $\omega = m\cosh\te$ is considered as an (unbounded)
multiplication operator on its maximal domain in
$L^2(\Rl,d\te)$. Along the same lines as in \cite[Prop. 2 (2)]{gl1},
one can show that these fields are localized on the left half line,
{\em i.e.} $\varphi(f)$, $\pi(f)$ are affiliated with $\A(W_L)$ if
$f\in\Ss(\Rl_-)$.\\
Choosing an $(n-1)$-particle vector $\xi_{n-1}\in\Hil_{n-1}$, an
operator $A\in\A(W_R)$, and a translation $s>0$, we define two linear functionals on $\Ss(\Rl)$ as
\begin{eqnarray}\label{def-cpm}
  \begin{array}{rcl}
  C^-_s(f) &:=& \langle\xi_{n-1},[\varphi(f),A(0,s)]\Omega\rangle,\\
  \vspace*{-3mm}\\
  C^+_s(f) &:=& \langle\xi_{n-1},[\pi(f),A(0,s)]\Omega\rangle\,.
  \end{array}
\end{eqnarray}
The operator $A$ is an arbitrary element of $\A(W_R)$, but fixed
in the following. The (anti-) linear dependence of the
above distributions on $A$ and $\xi_{n-1}$ will not be indicated in our notation.\\
We recall that the creation and annihilation operators $\zd(.),z(.)$ satisfy
the following bounds with respect to the particle number \cite{gl1},
familiar from free field theory ($\chi\in\Hil_1$):
\begin{eqnarray}\label{z-n-bnd}
  \begin{array}{rcl}
  \|z(\chi)\xi_{n-1}\|
  &\leq&
  (n-1)^{1/2}\|\chi\|\|\xi_{n-1}\|,\\
\vspace*{-2mm}\\
  \|\zd(\chi)\xi_{n-1}\|
  &\leq&
  n^{1/2}\|\chi\|\|\xi_{n-1}\|\,.
  \end{array}
\end{eqnarray}
Also note that for $f\in L^2(\Rl,dx)$ (we put $\|f\|_2 := (\int dx|f(x)|^2)^{1/2}$),
\begin{eqnarray}\label{of-2}
  \|\omega^{1/2}\fhat\|^2
  =
  \int d\te\, m\ch\te\,|\fhat(\te)|^2
  =
  \int dp\,|\fti(p)|^2
  =
  \|f\|_2^{\,2}\,.
\end{eqnarray}
Combining \bref{z-n-bnd} and \bref{of-2} with the Cauchy-Schwarz
inequality and using the annihilation property of $z(\cdot)$, we obtain the bounds
\begin{eqnarray}
  |C_s^\pm(\omega^{\mp 1/2}f)|
  \leq
  c_n \,\|\xi_{n-1}\|\|A\|\cdot\|f\|_2,\quad
  c_n &:=& \sqrt{n} + \sqrt{n-1} + 1\label{est-1}.
\end{eqnarray}
By carrying out the same estimate for $|C_s^\pm(f)|$ and taking into account
that both, $\|\fhat\|$ and $\|\omega\fhat\|$, can be dominated by
certain linear combinations of Schwartz space seminorms of $f$, we
first note $C_s^\pm \in \Ss'(\Rl)$. Moreover, \bref{est-1} shows that
the distributions $f\mapsto C_s^\pm(\omega^{\mp 1/2}f)$ are
regular in the sense that they are given by $L^2$-functions, whose
norm is bounded by $c_n\|\xi_{n-1}\|\|A\|$.\\

In view of the localization properties of $A$, $\varphi(f)$ and $\pi(f)$,
the support of $C_s^\pm$ is contained in the half line
$[s,\infty\,[\;\subset\Rl_+$. Consequently, their Fourier transforms
are boundary values (in the sense of distributions) of functions $\tilde{C}_s^\pm$ which are analytic in the lower half plane and polynomially bounded in imaginary direction
\cite[Thm. IX.16]{simon-reed-2}. Taking into account $\supp\,C_s^\pm
\subset [s,\infty\,[$, it follows that $\tilde{C}_s^\pm$ actually decays
exponentially in imaginary direction.\\

We now proceed to the rapidity picture by setting
\begin{eqnarray}\label{c-hat}
  \hat{C}_s^-(\te) := m\cosh(\te)\cdot\tilde{C}_s^-(m\sinh\te),
  \qquad
  \hat{C}_s^+(\te) := \tilde{C}_s^+(m\sinh\te)\,.
\end{eqnarray}
Note that because of the regularity of $\omega^{\mp
  1/2}\tilde{C}_s^\pm$, these are well-defined
{\em functions}. Their properties are collected in the following Lemma.

\begin{lemma}\label{lemma-C}
  Let $A\in\A(W_R)$, $s>0$, and $\xi_{n-1}\in\Hil_{n-1}$. The functions $\hat{C}^\pm_s$ \bref{c-hat} corresponding to the distributions \bref{def-cpm} have the following properties:
  \begin{itemize}
  \item[a)] $\hat{C}^\pm_s$ is the boundary value of a function analytic
    in the strip $S(-\pi,0)$.
 \item[b)] $\hat{C}_s^\pm$ is (anti-) symmetric with respect to reflections about $\Rl-\tfrac{i\pi}{2}$:
    \begin{eqnarray}\label{cpmsym}
        \hat{C}_s^\pm(\te-\tfrac{i\pi}{2}+i\mu)
        =
        \pm\hat{C}_s^\pm(-\te-\tfrac{i\pi}{2}-i\mu),\qquad -\tfrac{\pi}{2}<\mu<\tfrac{\pi}{2}\,.
    \end{eqnarray}
  \item[c)] Let $0\leq\la\leq\pi$ and $c_n = \sqrt{n}+\sqrt{n-1}+1$. The functions
    \begin{eqnarray}
      \Rl\ni\te\longmapsto \hat{C}^\pm_{s,\la}(\te) := \hat{C}^\pm_s(\te-i\la)
    \end{eqnarray}
    are elements of $L^2(\Rl,d\te)$, with norm bounded by
    $\|\hat{C}^\pm_{s,\la}\|\leq c_n\|\xi_{n-1}\|\|A\|$.
  \item[d)] $\overline{S(-\pi,0)}\ni\zeta\longmapsto \hat{C}^\pm_{s,\zeta}$ is continuous in the norm topology of $L^2(\Rl,d\te)$.
  \item[e)] For $\te\in\Rl$, $0<\la<\pi$,
    \begin{eqnarray}\label{c-bnd}
      |\hat{C}^\pm_s(\te-i\la)|
      &\leq&
        \sqrt{\frac{2}{\pi}}\,c_n\|\xi_{n-1}\|\|A\|\frac{e^{-\frac{ms}{2}\sin\la\,\cosh\te}}{\min\{\la,\pi-\la\}^{1/2}}\,.
    \end{eqnarray}
  \end{itemize}
\end{lemma}
\begin{proof}
a) Recalling
\begin{eqnarray}
  \begin{array}{rcl}
        \cosh(\te-i\la) &=& \cos\la\,\cosh\te-i\sin\la\,\sinh\te,\label{ch}\\
        \vspace*{-3mm}\\
        \sinh(\te-i\la) &=& \cos\la\,\sinh\te-i\sin\la\,\cosh\te,\label{sh}
  \end{array}
\end{eqnarray}
we see that $\sinh(.)$ maps the strip $S(-\pi,0)$ to the lower half plane. Hence $\hat{C}_s^\pm$ is analytic in $S(-\pi,0)$. b) is also a direct consequence of \bref{ch}.\\
To prove c), note that on the real line the claimed bound holds in view of the former estimates on $\|\omega^{\mp 1/2}\tilde{C}_s^\pm\|$:
\begin{eqnarray*}
  \int d\te\,|\hat{C}_s^\pm(\te)|^2
  =
  \int dp\,|\omega(p)^{\mp 1/2}\tilde{C}_s^\pm(p)|^2
  =
  \|\omega^{\mp 1/2}\tilde{C}_s^\pm\|_2^{\,2}
  \leq
  c_n^2\|\xi_{n-1}\|^2\|A\|^2.\label{bound-est1}
\end{eqnarray*}
But b) implies in particular $\hat{C}_s^\pm(\te-i\pi)=\pm
\hat{C}_s^\pm(-\te)$, and hence $\hat{C}_s^\pm$ is also square integrable
over the lower boundary $\Rl-i\pi$, with the same bound on its norm. As $\tilde{C}_s^\pm$ decays exponentially in imaginary direction, we have $\hat{C}_{s,\la}^\pm \in L^2(\Rl,d\te)$ for any fixed $0<\la<\pi$. Moreover, the limits $\lim_{\la\to 0}\hat{C}_{s,\la}^\pm$ and $\lim_{\la\to \pi}\hat{C}_{s,\la}^\pm$ are known to hold in the sense of distributions. These facts allow for the application of a version of the three lines theorem adapted to $L^2$-bounds, which is proven in the appendix as Lemma \ref{3lt-l2}. The results are: $\hat{C}_{s,\la}^\pm$ converges in the norm topology of $L^2(\Rl,d\te)$ as $\la\to 0$ or $\la\to \pi$, and the bound calculated on the boundary holds also for $\hat{C}_{s,\la}^\pm$, $0<\la<\pi$. This proves the claims c) and d).\\
Finally, e) is a consequence of a) and c): Let $\te\in\Rl$, $0<\la<\pi$, and put $\rho := \min\{\la,\pi-\la\}$. Then the disc $D_\rho$ with center
  $\te-i\la$ and radius $\rho$ is contained in the closed strip
  $\overline{S(-\pi,0)}$. By the mean value theorem for analytic
  functions, H\"older's inequality and the norm bound given in c), we get
\begin{eqnarray}
  |\hat{C}_s^\pm(\te-i\la)|
  &\leq&
  \frac{1}{\pi\rho^2}\int_{D_\rho}
  d\te'\,d\la'\,|\hat{C}_s^\pm(\te'+i\la')|\nonumber\\ 
  &\leq&
   \frac{1}{\sqrt{\pi}\rho}\left(\int_{D_\rho}
    d\te'\,d\la'\,|\hat{C}_s^\pm(\te'+i\la')|^2\right)^{1/2}\nonumber\\
 &\leq&
  \frac{1}{\sqrt{\pi}\rho}
  \left(\int_{-\la-\rho}^{-\la+\rho}d\la'\int_{-\infty}^\infty d\te'\,|\hat{C}_s^\pm(\te'+i\la')|^2\right)^{1/2}\nonumber\\
  &\leq&
  \sqrt{\frac{2}{\pi\rho}}\cdot c_n\|\xi_{n-1}\|\|A\|\,.\label{este-1}
\end{eqnarray}
Taking into account the covariance of $\varphi(f)$ and the translation
invariance of $\Omega$, one obtains
\begin{eqnarray*}
        C_s^-(f)
        &=&
        \langle \xi_{n-1},\,[\varphi(f),U(0,\tfrac{s}{2})A(0,\tfrac{s}{2})U(0,-\tfrac{s}{2})]\Omega\rangle\\
        &=&
        \langle U(0,-\tfrac{s}{2})\xi_{n-1},\,[\varphi(f_{\frac{s}{2}}),A(0,\tfrac{s}{2})]\Omega\rangle
        =:
        C'(f_{\frac{s}{2}}),
\end{eqnarray*}
where $f_{\frac{s}{2}}(x) = f(x+\frac{s}{2})$, $x\in \Rl$. Hence
$\hat{C}_s^-(\te) = e^{-\frac{ims}{2}\sh\te}\hat{C}'(\te)$, and
$\hat{C}'$ also fulfills the estimate \bref{este-1} since $\|U(0,\pm\frac{s}{2})\|=1$. Taking the absolute value of the exponential factor evaluated on points $\te-i\la\in S(-\pi,0)$ yields the claimed estimate \bref{c-bnd} for $\hat{C}_s^-$. The argument for $\hat{C}_s^+$ is the same.
\end{proof}

The relation between the functions $\hat{C}_s^\pm$ and the wavefunctions $\psi_n^s$ \bref{wave} is specified in the next Lemma.
\begin{lemma}\label{lemma-h}
Consider the function\footnote{We write $h^s_\xi$ instead of
  $h^s_{\xi_{n-1}}$ in order not to overburden our notation.}
\begin{eqnarray}\label{def-h}
   h^s_\xi(\zeta)
     &:=&  \frac{1}{2}(\hat{C}^-_s(\zeta)+i\hat{C}^+_s(\zeta)),
      \qquad
  \zeta\in\overline{S(-\pi,0)}\,,
\end{eqnarray}
which depends on $\xi_{n-1}\in\Hil_{n-1}$ through the definitions
\bref{def-cpm} and \bref{c-hat}. \\
  $h^s_\xi$ has the properties a), c), d), e) of the preceding Lemma. Moreover, $\te_1\in\Rl$,
\begin{eqnarray}
  h^s_\xi(\te_1) &=& \sqrt{n} \int d\te_2\cdots d\te_n\, \overline{\xi_{n-1}(\te_2,...,\te_n)}\cdot\psi_n^s(\te_1,...,\te_n)\,.\label{psi-h}
\end{eqnarray}
\end{lemma}
\begin{proof}
From the definition of $h^s_\xi$ it follows immediately that
properties a) and c)--e) of Lemma \ref{lemma-C} hold. To show \bref{psi-h}, let $f\in\Ss(\Rl)$.
  \begin{eqnarray}
    \langle \overline{\fhat_-},h^s_\xi\rangle
    &=&
    \frac{1}{2}\int d\te \fhat(-\te)
    \left(\hat{C}_s^-(\te)+i\hat{C}_s^+(\te)\right)\nonumber\\
    &=&
    \frac{1}{2}\int dp \fti(-p)
    \left(\tilde{C}_s^-(p)+i\omega(p)^{-1}\tilde{C}_s^+(p)\right)\nonumber\\
    &=&
    \frac{1}{2}\left(C_s^-(f)+iC_s^+(\omega^{-1}f)\right)\nonumber\\
    &=&
    \frac{1}{2}\langle\xi_{n-1},\,[\varphi(f)+i\pi(\omega^{-1}f),A(0,s)]\Omega\rangle\nonumber\\
    &=&
    \langle\xi_{n-1},\,[z(\fhat_-),A(0,s)]\Omega\rangle\label{alg1}\\
     &=&
    \langle\xi_{n-1},\,z(\fhat_-)A(0,s)\Omega\rangle\label{alg2}\\
    &=&
    \langle \zd(\overline{\fhat_-})\xi_{n-1},\,P_n U(0,s)A\Omega\rangle\label{alg3}\\
    &=&
    \sqrt{n}\;\langle
    \overline{\fhat_-}\otimes\xi_{n-1},\,\psi_n^s\rangle\,.\label{alg4}
  \end{eqnarray}
  In the last steps, we wrote the annihilation operator as a linear
  combination of the time zero fields in \bref{alg1}, {\em cf.}
  (\ref{def-phi},\ref{def-pi}), used the annihilation property of $z(.)$
  in \bref{alg2}, the relations of Zamolodchikov's algebra in
  \bref{alg3} and the definition of $\zd(.)$ in \bref{alg4}. By
  continuity, the above calculated relation
  \begin{eqnarray}
    \langle \overline{\fhat_-},h^s_\xi\rangle=\sqrt{n}\;\langle \overline{\fhat_-}\otimes\xi_{n-1},\,\psi_n^s\rangle
  \end{eqnarray}
  holds also if    
  $\fhat_-$ is replaced by an arbitrary function in $L^2(\Rl,d\te)$. This implies{\qquad} \bref{psi-h}.
\end{proof}

For $n=1$, Lemma \ref{lemma-h} states $\psi_1^s =
h^s_\Omega$. Hence the one particle wavefunctions $\psi_1^s$ have the
properties a), c)--e), listed in Lemma \ref{lemma-C}. In particular the
claims of Lemma \ref{lemma-n-rough} follow if the parameters $\alpha$
and $K$ appearing there are chosen as $\alpha = \frac{1}{4}$ and
$K=1$.\\
But for $n>1$, only information about $\psi_n^s$, considered as a function of the first variable $\te_1$, has been obtained. To extend this to the other variables, two properties of the scattering function \bref{s2gg},
\begin{eqnarray}
  S_2(\te) &=& \prod_{r=1}^R \frac{1+ig_r\sinh\te}{1-ig_r\sinh\te},\qquad g_1,...,g_R>0,
\end{eqnarray}
are important, which we extract now. Firstly, we see that $S_2$ is a meromorphic function in the entire
complex plane, with certain $(2\pi i)$-periodic sequences of poles. In
particular, $S_2$ is analytic not only in the physical sheet $S(0,\pi)$, but also in the
wider strip $S(-\kappa_g,\pi+\kappa_g)$, where $\kappa_g>0$ is given by 
\begin{eqnarray}\label{s-kappa}
  \kappa_g := \left\{
    \begin{array}{rcl}
      \arcsin(g^{-1}) & ;\quad & g>1\\
      \frac{\pi}{2}           & ;\quad & 0<g\leq 1
    \end{array}
  \right.,\qquad\quad g := \max_{r=1,...,R}g_r\,.
\end{eqnarray}
Furthermore, given arbitrary $\kappa\in[\,0,\kappa_g\,[$, we note that $S_2$ is
uniformly bounded on $S(-\kappa, \pi+\kappa)$. This bound will be
denoted by
\begin{eqnarray}\label{uniform-s-bound}
  \sigma(\kappa)
  :=
  \sup\{|S_2(\zeta)|\,:\, \zeta\in\overline{S(-\kappa,\pi+\kappa)}\}
  < \infty,\qquad 0<\kappa < \kappa_g\,.
\end{eqnarray}
These two features of $S_2$, the analyticity in the enlarged strip and the bound \bref{uniform-s-bound}, will be essential in the following proof of Lemma \ref{lemma-n-rough}. Furthermore, the parameters $\alpha$ and $K$ appearing there, will be specified explicitely in terms of $\kappa,\sigma(\kappa)$ and $n$.\\

\noindent{\bf Proof of Lemma \ref{lemma-n-rough}:}\\
Fix $\zeta_1=\te_1-i\la_1\in S(-\pi,0)$. In view of the definitions of $C_s^\pm$ \bref{def-cpm}, $\hat{C}_s^\pm$ \bref{c-hat} and $h^s_\xi$ \bref{def-h}, it is clear that $\xi \longmapsto h^s_\xi(\zeta_1)$ is an anti-linear functional on $\Hil_{n-1}$, which  by Lemma \ref{lemma-C} e) is norm-continuous. Hence, by application of Riesz' Theorem, $h^s_\xi(\zeta_1)$ can be written as
\begin{eqnarray}\label{rel-xi-psi}
  h^s_\xi(\zeta_1)
  &=:&
  \sqrt{n}\int d\te_2\cdots d\te_n\,
  \overline{\xi_{n-1}(\te_2,...,\te_n)}\cdot \psi_n^s(\zeta_1,\te_2,...,\te_n)\,,
\end{eqnarray}
where $(\te_2,...,\te_n)\mapsto\psi_n^s(\zeta_1,\te_2,...,\te_n)$ is a
function in $\Hil_{n-1}\subset L^2(\Rl^{n-1}d^{n-1}\bte)$
which is defined by this equation. Taking $\xi_{n-1}$ to be the
characteristic function of a compact set
$\mathcal{C}\subset\Rl^{n-1}$, and $\chi_I$ to be the characteristic
function of a compact interval $I\subset\Rl$, it follows from
\begin{eqnarray*}
 \sqrt{n}\left|\,\int\limits_{I\times\mathcal{C}} d^n\bte\,
  \psi_n^s(\te_1-i\la_1,\te_2,...,\te_n)\right|
  &=&
  \left|\int d\te_1 \overline{\chi_I(\te_1)} h_{\xi}^s(\te_1-i\la_1)\right|
  <\infty
\end{eqnarray*}
 that $\bte\longmapsto\psi_n^s(\te_1-i\la_1,\te_2,...,\te_n) =:
 \psi_{n,\la_1}^s(\bte)$ is locally integrable and hence
 measurable. According to Lemma \ref{lemma-C} e),
\begin{eqnarray}\label{bbbnd}
        \int d\te_2\cdots d\te_n\,|\psi_{n,\la_1}^s(\bte)|^2
                &\leq& 
        \frac{2c_n^2\|A\|^2 \,e^{-ms\sin\!\la_1\ch\te_1}}{\pi n \min\{\la_1,\pi-\la_1\}}.
\end{eqnarray}
In view of the measurability of $\psi_{n,\la_1}^s$, it follows from the estimate \bref{bbbnd} that for fixed $\la_1\in(0,\pi)$,
$\psi_{n,\la_1}^s$ is an element of
$L^2(\Rl^n,d^n\bte)$. Furthermore, note that the set 
$\{n^{-1}c_n^2\,:\,n\in\N\}$ is bounded. Hence there is a function
$b(\la_1,s)$, independent of $n$ and $A$, such that
\begin{eqnarray}\label{b-bnd}
  \|\psi_{n,\la_1}^s\| &\leq& b(\la_1,s)\cdot\|A\|\,,\qquad 0<\la_1<\pi,\;s>0\,.
\end{eqnarray}
Next we want to show that $\psi_{n,\la_1}^s$ converges to $\psi_n^s$ in the
norm topology of $L^2(\Rl^n,d^n\bte)$ as $\la_1\to 0$. To this
end, we need to improve on the bound \bref{b-bnd}, because
$b(\la_1,s)$ diverges for $\la_1\to 0$ and for $\la_1\to \pi$. Note
first that as a consequence of its rapid decrease in the real
direction, the Fourier transform of $(h^s_\xi)_{\la_1}$ is given by
$\beta_1 \mapsto e^{\la_1\beta_1}\widetilde{h^s_\xi}(\beta_1)$, $\beta_1\in\Rl$ ({\em
  cf.} the proof of Lemma \ref{3lt-l2} in the appendix). In view of
\bref{rel-xi-psi}, this relation implies that the Fourier transform of
$\psi_{n,\la_1}^s$ is $\Rl^n\ni\bbe\mapsto e^{\la_1\beta_1}\cdot\widetilde{\psi_n^s}(\bbe)$. Now one can proceed as in
the proof of Lemma \ref{3lt-l2} and apply Lebesgue's dominated
convergence theorem to arrive at the norm limit
$\psi_{n,\la_1}^s\to\psi_n^s$ as $\la_1\to 0$. The limit $\psi_{n,\la_1}^s \to \psi_{n,\pi}^s$ as $\la_1\to \pi$
can be established by exploiting the symmetry properties (Lemma
\ref{lemma-C} b)) of $\hat{C}_s^\pm$, but will not be needed here.\\
We now turn to the analyticity properties of the wavefunctions and
consider a point $\zeta_0\in S(-\pi,0)$ with an appropriate curve
$\mathcal{C}_0\subset S(-\pi,0)$ containing it in its interiour. As $h^s_\xi$ is analytic, we have
\begin{eqnarray*}
  0
  =
  \oint_{\mathcal{C}_0}d\zeta\, h^s_\xi(\zeta) 
  =
  \sqrt{n}\oint_{\mathcal{C}_0}d\zeta\, \int d\te_2\cdots d\te_n\,\overline{\xi_{n-1}(\te_2,...,\te_n)}\,\psi_n^s(\zeta,\te_2,...,\te_n).
\end{eqnarray*}
But since the integrand is integrable over $\mathcal{C}_0 \times \Rl^{n-1}$, we may reverse the order of the two
integrals and conclude $\oint_{\mathcal{C}_0} d\zeta\,
\psi_n^s(\zeta,\te_2,...,\te_n) = 0$ for almost all
$\te_2,...,\te_n\in\Rl$, {\em i.e.} $\psi_n^s$ is the boundary value
of an analytic function in the first variable if the other variables
$\te_2,...,\te_n\in\Rl$ are held fixed.\\
Consider the symmetry condition ({\em cf.} \bref{psi-sym})
\begin{eqnarray}\label{sym1}
  \psi_n^s(\te_2,\te_1,\te_3,...,\te_n)
  &=&
  S_2(\te_1-\te_2)\cdot\psi_n^s(\te_1,\te_2,\te_3,...,\te_n)
\end{eqnarray}
and let $\te_2,...,\te_n\in\Rl$ be fixed. In view of the analyticity
of the scattering function in the enlarged strip
$S(-\kappa_g,\pi+\kappa_g)$, we see that the right hand side of
\bref{sym1} is the boundary value of a function analytic in its first
variable $\te_1$, in the region $S(-\kappa_g,\pi+\kappa_g) \cap S(-\pi,0) = S(-\kappa_g,0)$. Hence the left hand side can be continued to $S(-\kappa_g,0)$ as well, and we conclude that $\psi_n^s$ has also an analytic extension in the second variable, to the strip $S(-\kappa_g,0)$, if the other variables are held fixed. In the same way we see inductively from 
\begin{eqnarray}\label{sym2}
 \!\!\!\! \psi_n^s(\te_1,...,\te_{k+1},\te_k,...,\te_n)
  &=&
  S_2(\te_k-\te_{k+1})\psi_n^s(\te_1,...,\te_k,\te_{k+1},...,\te_n)
\end{eqnarray}
that $\psi_n^s$ is analytic in each variable $\te_k\in S(-\kappa_g,0)$ if
the remaining $(n-1)$ variables are held fixed and real. (We neglect here the
even stronger analyticity in the first variable in the region $S(-\pi,0)$.) By application of the
Malgrange-Zerner theorem (for a proof of this theorem see \cite{epstein}) we
conclude that $\psi_n^s$ is analytic (as a function of $n$ complex
variables) in the tube region
\begin{eqnarray}
  \Tu_n(\kappa_g)
  &:=&
  \Rl^n - i\mathscr{B}_n(\kappa_g),\label{tube}\\
  \mathscr{B}_n(\kappa_g)
  &:=& 
  \bigg\{\bla\in \;]\,0,\kappa_g\,[^{\times n}\,:\;0<\sum_{k=1}^n\la_k<\kappa_g\bigg\}\,.\label{base}
\end{eqnarray}
To avoid the divergencies due to the poles of $S_2$ at the boundary of
$\Tu_n(\kappa_g)$, we now fix some $\kappa\in\;]\,0,\kappa_g]$ and
consider the smaller tube $\Tu_n(\kappa)$ instead of $\Tu_n(\kappa_g)$. 
Note that $\mathscr{B}_n(\kappa)$ contains the $n$-dimensional cube $]\,0,\frac{\kappa}{n}\,[^{\times n}$. As
\begin{eqnarray}
  (\Delta^\la\psi_n^s)(\bte) 
  &=&
  \psi_n^s(\te_1-2\pi\la\cdot i,...,\te_n-2\pi\la\cdot i)\,,
\end{eqnarray}
the analyticity of $\psi_n^s$ in $\Tu_n(\kappa)$ implies the claim of Lemma \ref{lemma-n-rough} a), and the parameter $\alpha$ appearing there can be chosen as
\begin{eqnarray}
  \alpha = \frac{\kappa}{4\pi n}\,.
\end{eqnarray}
Now we use the uniform bound \bref{uniform-s-bound} to prove part b). The relations
\bref{sym1} and \bref{b-bnd} imply $\|\psi_n^s(\,.\,,\,.-i\la_2,...)\|\leq \sigma(\la_2)b(\la_2,s)\|A\|$, and
inductively we get from \bref{sym2}
\begin{eqnarray}
  \int d^n\bte\, |\psi_n^s(\te_1,...,\te_k-i\la_k,...,\te_n)|^2
  &\leq& 
  \sigma(\la_k)^{2(k-1)}b(\la_k,s)^2\cdot\|A\|^2.
\end{eqnarray}
Thus the analytic
continuations of $\psi_n^s$ in each single variable have $L^2$-norm bounded
by $\sigma(\kappa)^{n-1}b(\kappa,s)\|A\|$. This bound can be
transported to the interiour of the tube $\Tu_n(\kappa)$ by using the
flat tube theorem, see Lemma \ref{flattube-bounds} in the
appendix. We arrive at
\begin{eqnarray}\label{normbound}
  \int d^n\bte\, |\psi_n^s(\bte-i\bla)|^2
  &\leq& 
  \sigma(\kappa)^{2(n-1)}b(\kappa,s)^2\cdot\|A\|^2\,,\quad
        \bla\in\overline{\mathscr{B}_n(\kappa)}\,.
\end{eqnarray}
In particular, the norm bound claimed in Lemma \ref{lemma-n-rough} b) follows, and the parameter $K$ can be chosen as
\begin{eqnarray}
  K = \sigma(\kappa)^{n-1}\cdot b(\kappa,s)\,.
\end{eqnarray}
To establish the limit $\lim_{|\sbte|\to\infty}\psi_n^s(\bte-i\bla)=0$, let $\bte\in\Rl^n$, $\bla\in\mathscr{B}_n(\kappa)$ and consider a
polydisc $\mathscr{D}_\rho\subset\Tu_n(\kappa)$ with sufficiently small radius $\rho$ and $\bte-i\bla\in\mathscr{D}_\rho$. By the mean value property for analytic functions and H\"older's inequality,
  \begin{eqnarray*}
        |\psi_n^s(\bte-i\bla)|^2
        &\leq&
        (\pi\rho^2)^{-n}\int\limits_{\mathscr{D}_\rho} d^n\bze'\,|\psi_n^s(\bze')|^2\\
        &\leq& 
        (\pi\rho^2)^{-n} \int\limits_{[-\rho,\rho]^{\times n}}d^n
        \bla'
        \int\limits_{[-\rho,\rho]^{\times n}}d^n\bte'
        |\psi_n^s(\bte'+\bte+i(\bla'-\bla))|^2\,.
  \end{eqnarray*}
Because of \bref{normbound} the last integral is convergent, also when the integration in $\bte'$ is carried out over $\Rl^n$ instead of $[-\rho,\rho]^{\times n}$. Hence it vanishes in the limit $|\bte|\to\infty$.\\
Finally, c) is a consequence of the earlier discussed strong continuity (Lemma \ref{lemma-C} d)) of
  $[0,\kappa]\ni\la_1\mapsto\psi_{n,\la_1}^s$. {\hfill $\square$ \\[2mm] \indent}

\section{Summary and Outlook}

We have reviewed a novel approach to the construction of quantum field
theories with a factorizing S-matrix on two-dimensional Minkowski
space. Starting from a scattering function describing the interaction
of one type of massive, scalar particles without bound states, in a
first step certain auxiliary quantum fields (``polarization-free generators'') were constructed. It was shown how to define a covariant net of wedge
algebras from these fields. Furthermore, we mentioned that the
``correct'' two-particle scattering behaviour, namely the one expected
from the input scattering function, can be recovered from the
wedge-local fields.\\
In a second step, the local operator content of
these wedge local theories has to be analyzed to ensure physical
meaningful models. In this context, the modular nuclearity condition
constitutes a sufficient criterion for the existence of local observables. In
two particular examples, namely the models with constant scattering
functions $\pm 1$, this criterion has been verified already.\\
In
the present paper, the modular compactness criterion, a
necessary condition for the split property of the wedge net and thus providing an intermediate step in proving the existence of local operators, has been
checked in a wide class of models with typical, non-constant
scattering functions. In view of these results, it seems reasonable to
conjecture that the question of the existence of local observables
will have an affirmative answer in the family of models considered.\\
If this conjecture can be proven, the
program reviewed here provides a possibility to rigorously construct
interacting quantum field theories in two dimensions, without taking
recourse to classical concepts. Although the family of S-matrices considered is limited, it seems possible to generalize the
procedure to more complicated models with several kinds of massive
particles, ultimately leading to an existence proof for quantum
field theories with arbitrary factorizing S-matrix.

\section*{Appendix}
\renewcommand{\theequation}{A.\arabic{equation}}
\renewcommand{\thetheorem}{A.\arabic{theorem}}

In this appendix we prove two Lemmata in complex analysis which are
used in the main text. The first one is an adaptation of the three
lines Theorem \cite[Thm 12.8]{rudin} to the case of $L^2$-bounds, and the second one shows how to obtain such bounds in the situation of the Malgrange-Zerner theorem. Both statements seem to be well-known; but as we did not find them in this form in the literature, we give their proofs here.

\begin{lemma}\label{3lt-l2}
  Let $a,b\in\Rl$, $a<b$,
  \begin{eqnarray}
    S(a,b) &:=& \{z\in\Cl\,:\,a<\mathrm{Im}(z)<b\},
  \end{eqnarray}
  and let $f$ denote a function which is analytic in $S(a,b)$. Assume
  that for each $y\in[a,b]$, the function $x\mapsto f(x+iy) =: f_y(x)$
  is an element of $L^2(\Rl,dx)$ and that $f_y \to f_c$ for $y\to c$,
  where $c=a,b$, in the sense of distributions.\\
  Then the limits $f_y\to f_c$ are also valid in the norm topology of
  $L^2(\Rl,dx)$, and 
  \begin{eqnarray}
    \|f_y\|
    &\leq&
    \max\{\|f_a\|,\|f_b\|\}\,,\qquad a\leq y \leq b\,.
  \end{eqnarray}
\end{lemma}
\begin{proof}
  Let $\gti\in C_0^\infty(\Rl)$. Then $g$ is entire analytic and
  $x\mapsto g(x+iy) =: g_y(x)$ is of rapid decrease at infinity for fixed
  $y\in\Rl$. Let $0<y<b-a$.
  \begin{eqnarray*}
    \int dp\,\fti_{a+y}(-p)\gti(p)
    &=&
    \int dx\, f_{a+y}(x)g(x)
    =
    \int dx\,f_{a+\eps y}(x)g_{(\eps-1)y}(x).
  \end{eqnarray*}
  Here we used the rapid decrease of $g_y$ and the analyticity of $f$
  and $g$ to shift the integration from $\Rl$ to $\Rl+iy(\eps-1)$,
  where $0<\eps<1$. As $\gti\in C_0^\infty(\Rl)$, the limit
  $\lim_{\eps\to 0}g_{(\eps-1)y} = g_{-y}$ holds in the topology of $\Ss(\Rl)$. Together
  with the distributional convergence $f_{a+\eps y}\to f_a$, this
  implies that the above integral is equal to
  \begin{eqnarray*}
    \int dx\, f_a(x)g(x-iy)
    &=& 
    \int dp\, \fti_a(-p) e^{yp}\gti(p)\,.
  \end{eqnarray*}
  Hence $\fti_{a+y}(p)=e^{-yp}\fti_a(p)$. This implies
  \begin{eqnarray*}
    \|f_{a+y} - f_a\|^2
    &=&
    \int dp\,|\fti_a(p)|^2(e^{-yp}-1)^2\,,
  \end{eqnarray*}
  and since we have the integrable bound
  \begin{eqnarray*}
    |\fti_a(p)|^2(e^{-yp}-1)^2
    &\leq&
    |\fti_a(p)|^2\left(4\Theta(p)+\Theta(-p)(1+e^{-p(b-a)})^2\right),
  \end{eqnarray*}
  we may use Lebesgue's dominated convergence theorem to conclude
  $\lim_{y\searrow a}f_y = f_a$ in the norm toplogy of $L^2(\Rl,dx)$. The
  limit $\lim_{y\nearrow b}f_y = f_b$ is established in the same manner.\\
  Now let $h\in\Ss(\Rl)$ be a test function and consider the
  convolution $f * h$, which is an analytic function in
  $S(a,b)$. It satisfies the bound, $a<y<b$, 
  \begin{eqnarray*}    
    |(f*h)(x+iy)|
    &\leq&
    \int dx'\, |h(x')|\cdot |f_y(x-x')|
    \leq
    \|h\|\cdot\|f_y\|
    <
    \infty\,.
  \end{eqnarray*}
  But in view of the above established continuity of $[a,b]\ni y\mapsto
  f_y$, the norm $\|f_y\|$ depends continuously on $y$, and
  hence we can find a uniform bound on $|(f*h)(z)|$, $z\in
  \overline{S(a,b)}$. By the three lines theorem, we conclude
  \begin{eqnarray*}
    \left|\int dx'\, h(x') f_y(x-x')\right|
    &\leq&
    \|h\|\cdot\max\{\|f_a\|, \|f_b\|\}
    ,\qquad a\leq y \leq b\,.
  \end{eqnarray*}
  As $h\in\Ss(\Rl)$ was arbitrary and $\Ss(\Rl)$ is dense in
  $L^2(\Rl,dx)$, the claim follows.
\end{proof}

\begin{lemma}\label{flattube-bounds}
  Let
  \begin{eqnarray}
    \mathscr{B}_n
    :=
    \bigg\{\by\in (0,1)^{\times n} \,:\, 0<\sum_{j=1}^n y_j<1\bigg\},\qquad
      \Tu_n
      :=
      \Rl^n + i\mathscr{B}_n
    \end{eqnarray}
    and consider an analytic function $f : \Tu_n\to\Cl$ of $n$ complex
    variables. Setting $f_{\sby} : \Rl^n\to\Cl, \bx\mapsto
    f(\bx+i\by)$, assume that $f_{\sby}\in L^2(\Rl^n,d^n\bx)$
    for any $\by\in\mathscr{B}_n$, that the map $\by\mapsto f_{\sby}$
    can be extended norm continuously to $\overline{\Tu_n}$, and that
    \begin{eqnarray}
      \|f_{(0,...,y_k,...0)}\|_2 &\leq& 1,\qquad 0<y_k<1,\;k\in\{1,...,n\},
    \end{eqnarray}
    holds.\\
    Then one has the bound
    \begin{eqnarray}\label{lemma-claim}
      \|f_{\sby}\|_2 &\leq& 1 
    \end{eqnarray}
    for any $\by\in \overline{\mathscr{B}_n}$.
  \end{lemma}
\begin{proof}
  Let $g\in\Ss(\Rl^n)$ be a test function and consider the convolution
  $f*g$. According to the hypothesis of the Lemma, $f*g$ is analytic in
  $\Tu_n$ and $\by\longmapsto {(f*g)}_{\sby}$ is continuous on
  $\overline{\Tu_n}$. On the boundary we have
  \begin{eqnarray}\label{lemma-bound}
    |(f*g)(x_1,...,x_k+iy_k,...,x_n)|
    &\leq&
    \|g\|_2\|f_{(0,...,y_k,...0)}\|_2
    \;\leq\;
    \|g\|_2.
  \end{eqnarray}
  Now consider $h_g(\bz) := ((f*g)(\bz)-e^{i\alpha}(\|g\|+\eps))^{-1}$,
  where $\eps>0$ and $\alpha\in\Rl$ are arbitrary. Due to the bound
  \bref{lemma-bound}, $h_g$ is, in each variable seperately, analytic
  in the strip $S(0,1)$ if the remaining variables are held fixed and
  real. By the Malgrange-Zerner Theorem, we conclude that $h_g$ has an
  analytic continuation, as a function of $n$ complex variables, to
  the tube $\Tu_n$. Varying $\alpha$ and letting $\eps\to 0$, we
  conclude 
  \begin{eqnarray}
    |(f*g)(\bx+i\by)|
    &=& |\langle \hat{g}_{\sbx},\, f_{\sby}\rangle |
    \;\leq\;
    \|g\|_2
    \;=\;
    \|\hat{g}_{\sbx}\|_2 \,,
  \end{eqnarray}
where we have put $\hat{g}_{\sbx}(\bx') :=
\overline{g(\bx-\bx')}$. But as $g\in\Ss(\Rl^n)$ was arbitrary
and $\Ss(\Rl^n)$ is norm dense in
$L^2(\Rl^n,d^n\bx)$, the claim \bref{lemma-claim} follows.
\end{proof}

\subsection*{Acknowledgements}
Many thanks are due to D.~Buchholz for supervision and advice. I also
benefitted from conversations with K.-H.~Rehren and
B.~Schroer. Regarding the theory of complex analysis, several
discussions with J.~Bros and H.-J.~Borchers have been very helpful. Financial support by the DFG (Deutsche Forschungsgemeinschaft) is acknowledged.



\begin{thebibliography}{99}
\footnotesize

\bibitem{araki} H.~Araki: {\it Mathematical Theory of Quantum Fields},
  Oxford University Press, New York (1999)

\bibitem{ffp} H.~Babujian, M.~Karowski: {\it The "Bootstrap Program"
    for Integrable Quantum Field Theories in 1+1 Dim}, preprint,
  [arXiv: hep-th/0110261] (2001)

\bibitem{BKW} B.~Berg, M.~Karowski, P.~Weisz: {\it Construction of
    Green's functions from an exact $S$ matrix}, Phys. Rev. D {\bf
    19}, 2477 (1979)

\bibitem{BiWi} J.J.~Bisognano, E.H.~Wichmann: {\it On the duality
    condition for a hermitian scalar field}, J. Math. Phys. {\bf 16}, 
  985 (1975)

\bibitem{borchers} H.-J.~Borchers: {\it The CPT theorem in two-dimensional
    theories of local observables}, Commun. Math. Phys. {\bf 143}, 315 (1992)

\bibitem{pfgs}  H.-J.~Borchers, D.~Buchholz, B.~Schroer:
  {\it Polarization-Free Generators and the S-Matrix},
  Commun. Math. Phys. {\bf 219}, 125 (2001) [arXiv: hep-th/0003243]

\bibitem{bros:compactness} J.~Bros: {\it A Proof of Haag-Swieca's
    Compactness Property for Elastic Scattering States},
  Commun. Math. Phys {\bf 237}, 289 (2003)

\bibitem{BGL} R.~Brunetti, D.~Guido, R.~Longo: {\it Modular
    localization and Wigner particles}, Rev. Math. Phys. {\bf 14},
  759 (2002), [arXiv: math-ph/0203021]

\bibitem{bu:productstates} D.~Buchholz: {\it Product States for Local
    Algebras}, Commun. Math. Phys {\bf 36}, 287 (1974)

\bibitem{nuclearmaps1} D.~Buchholz, C.~D'Antoni, R.~Longo {\it Nuclear
    Maps and Modular Structures I: General Properties},
  Journ. Funct. Anal. {\bf 88}, 233 (1990)

\bibitem{nuclearmaps2} D.~Buchholz, C.~D'Antoni, R.~Longo {\it Nuclear
    Maps and Modular Structures II: Applications to Quantum Field Theory},
  Commun. Math. Phys. {\bf 129}, 115 (1990)

\bibitem{BuJa} D.~Buchholz, P.~Jacobi: {\it On the nuclearity condition
    for massless fields}, Lett. Math. Phys. {\bf 13}, 313 (1987)

\bibitem{BuJu} D.~Buchholz, P.~Junglas: {\it On The Existence of
    Equilibrium States in Local Quantum Field Theory}, Commun. Math. Phys. {\bf 121}, 255 (1989)

\bibitem{BuLe} D.~Buchholz, G.~Lechner: {\it Modular Nuclearity and
    Localization}, Ann. H. Poincar\'e {\bf 5}, 1065 (2004) [arXiv: math-ph/0402072]

\bibitem{BuWi} D.~Buchholz, E.~H.~Wichmann: {\it Causal Independence
    and the Energy-Level Density of States in Local Quantum Field
    Theory}, Comm. Math. Phys. {\bf 106}, 321 (1986) 

\bibitem{castro} O.~A.~Castro-Alvaredo: {\it Bootstrap Methods in
    1+1-Dimensional Quantum Field Theories: The Homogeneous
    Sine-Gordon Models}, PhD thesis (2001) [arXiv: hep-th/0109212]

\bibitem{dopl} S.~Doplicher, R.~Longo: {\it Standard and split
    inclusions of von Neumann algebras}, Commun. Math. Phys {\bf 75},
  493 (1984)

\bibitem{epstein} H.~Epstein: {\it Some analytic properties of
    scattering amplitudes in quantum field theory}, in: {\it
    Particle Symmetries and Axiomatic Field Theory}, Brandeis Summer
  School 195, New York: Gordon and Breach (1966)

\bibitem{haag} R.~Haag: {\it Local Quantum Physics}, Berlin: Springer
  Verlag, second edition (1996)

\bibitem{haag-swieca} R.~Haag, J.~A.~Swieca: {\it When does a quantum
    field theory describe particles?}, Commun. Math. Phys. {\bf 1},
  308 (1965)

\bibitem{kadring} R.~V.~Kadison, J.~R.~Ringrose: {\it Fundamentals of
  the Theory of Operator Algebras}, Vol. II, Orlando: Academic Press (1986)

\bibitem{gl1} G.~Lechner: {\it Polarization-Free Quantum Fields and
    Interaction}, Lett. Math. Phys. {\bf 64}, 137 (2003), [arXiv: hep-th/0303062]

\bibitem{gl2} G.~Lechner: {\it On the existence of local observables
    in theories with a factorizing S-matrix}, to appear in Journ. of
  Phys. {\bf A} (2005) [arXiv: math-ph/0405062]

\bibitem{ligmin} A.~Liguori, M.~Mintchev: {\it Fock spaces with
    generalized statistics}, Commun. Math. Phys. {\bf 169}, 635 (1995),
  [arXiv: hep-th/9403039]

\bibitem{longo} R.~Longo: {\it Notes on algebraic invariants for
    noncommutative dynamical systems}, Commun. Math. Phys. {\bf 69},
  195 (1979)

\bibitem{mccoy} B.~M.~McCoy, C.~A.~Tracy, T.~T.~Wu: {\it Two
    Dimensional Ising Model as an Exactly Solvable Relativistic
    Quantum Field Theory: Explicit Formulas for $n$-Point Functions} Phys. Rev. Lett. {\bf 38}, 793 (1977)

\bibitem{mitra} P.~Mitra: {\it Elasticity, Factorization and
    S-Matrices in (1+1)-Dimensions}, Phys. Lett. {\bf 72 B}, 62 (1977)

\bibitem{mueger:spw} M.~M\"uger: {\it Superselection structure of
    massive quantum field theories in (1+1)-dimensions.}
  Rev. Math. Phys. {\bf 10}, 1147 (1998) [arXiv: hep-th/9705019]

\bibitem{mund} J. Mund: {\it The Bisognano-Wichmann theorem for
    massive theories}, Annales Henri Poincar\'e {\bf 2}, 907 (2001), 
  [arXiv: hep-th/0101227]

\bibitem{pietsch} A.~Pietsch: {\it Nuclear locally convex spaces}
  Berlin, Heidelberg, New York: Springer (1972)

\bibitem{rudin} W.~Rudin: {\it Real and complex analysis}, McGraw-Hill
  Book Company (1987)

\bibitem{sakai} S.~Sakai: {\it $C^*$-Algebras and $W^*$-Algebras},
  Springer (1971)

\bibitem{schroer1} B.~Schroer: {\it Modular localization and the
    bootstrap form-factor program}, Nucl.Phys. {\bf B 499},
547 (1997), [arXiv: hep-th/9702145]

\bibitem{schroer-rqft} B.~Schroer: {\it New Constructions in Local
    Quantum Physics}, to appear in proceedings of the conference
  ``Rigorous Quantum Field Theory'', Saclay 2004 [arXiv: hep-th/0501104]

\bibitem{schroer-wiesbrock} B.~Schroer, H.~W.~Wiesbrock: {\it Modular
    constructions of quantum field theories with interactions}, 
  Rev. Math. Phys {\bf 12}, 301 (2000), [arXiv: hep-th/9812251]

\bibitem{smirnov} F.~A.~Smirnov: {\it Formfactors in completely
    integrable models in quantum field theory}, Advanced Series in
  Mathematical Physics 14, World Scientific (1992)

\bibitem{simon-reed-1} M.~Reed, B.~Simon: {\it Methods of modern
    mathematical physics I: Functional Analysis}, Revised and enlarged
  edition, Academic Press (1980)

\bibitem{simon-reed-2} M.~Reed, B.~Simon: {\it Methods of modern
    mathematical physics II: Fourier Analysis, Self-Adjointness},
  Academic Press (1975)

\bibitem{simon-reed-3} M.~Reed, B.~Simon: {\it Methods of modern
    mathematical physics III: Scattering Theory},
  Academic Press (1979)

\bibitem{summers:statistical} S.~J.~Summers: {\it On the Independence of Local Algebras in Quantum Field Theory}, Rev. Math. Phys. {\bf 2}, 201 (1990)

\bibitem{zamo} A.~Zamolodchikov: {\it Factorized S-matrices as the
    exact solutions of certain relativistic quantum field theory
    models}, Ann. Phys. {\bf 120}, 253 (1979)

\end{thebibliography}
\end{document}